\documentclass[aip,reprint]{revtex4-1}

\usepackage{graphicx}
\usepackage{amsmath}
\usepackage{amssymb}

\usepackage{color}

\draft 

\begin{document}
%
%
%
%
%
%


\title{How does Spreading Depression Spread? - Physiology and Modeling}

\author{Bas-Jan Zandt}
\email[]{Corresponding author:\\
Bas-Jan Zandt, PhD\\
Email: Bas-Jan.Zandt@biomed.uib.no\\ 
Department of Biomedicine\\
Postboks 7800\\ 
5020 Bergen\\
Norway
}
\affiliation{Department of Biomedicine, University of Bergen, Bergen, Norway}

\author{Bennie ten Haken}

\affiliation{MIRA-Institute for Biomedical Technology and Technical Medicine, University of Twente, Enschede, the Netherlands}
\author{Michel J.A.M. van Putten}

\affiliation{MIRA-Institute for Biomedical Technology and Technical Medicine, University of Twente, Enschede, the Netherlands}
\affiliation{Department of Clinical Neurophysiology, Medisch Spectrum Twente, Enschede, the Netherlands}

\author{Markus A. Dahlem}

\affiliation{Department of Physics, Humboldt-Universit\"at zu Berlin, Berlin, Germany.}

\begin{abstract}
Spreading depression (SD) is a wave phenomenon in gray matter tissue. Locally, it is characterized by massive re-distribution of ions across cell membranes. As a consequence, there is a sustained membrane depolarization and tissue polarization that depresses any normal electrical activity.  Despite these dramatic cortical events, SD remains difficult to observe in humans noninvasively, which for long has slowed advances in this field.  The growing appreciation of its clinical importance in migraine and stroke is therefore consistent with an increasing need for computational methods that tackle the complexity of the problem at multiple levels. In this review, we focus on mathematical tools to investigate the question of spread and its two complementary aspects: What are the physiological mechanisms and what is the spatial extent of SD in the cortex? This review discusses two types of models used to study these two questions, namely Hodgkin-Huxley type and generic activator-inhibitor models, and the recent advances in techniques to link them.
\end{abstract}


\maketitle

\section{Introduction and Scope}
\begin{figure}
\label{fig:KandV}
		\includegraphics[width=1.0\linewidth]{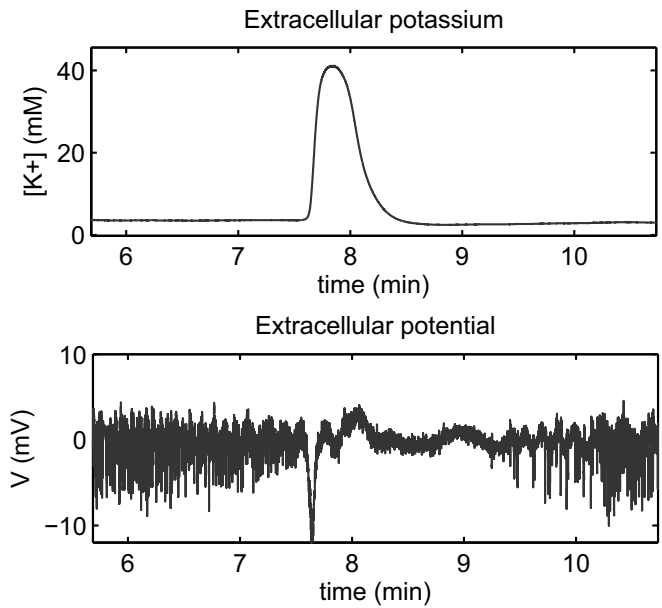}
		\caption{Disturbed ion concentrations and suppression of neuronal electrical activity. Extracellular potassium concentration (upper panel) and extracellular potential (lower panel) during SD at a fixed position (in vivo rat cortex). Sudden release of neuronal K$^+$ into the extracellular space is observed at t = 7.7 min. Redistribution of ions (K$^+$, Na$^+$, Cl$^-$ and Ca$^{++}$), between the intra- and extracellular space results in temporary neuronal dysfunction and cessation of extracellular electrical activity. In these normoxic conditions, recovery of [K$^+$] (after 1 min) and electrical activity (after 2-3 min) is relatively fast. The disturbance was observed to be traveling over the cortex at several mm/min. (Backes, Feuerstein, Ima, Zandt and Graf, unpublished data)}
\end{figure}

Spreading depression (SD), or depolarization\footnote{The term spreading depolarization is sometimes used to distinguish between the wave in the tissue, and the observed depression of neuronal electrical activity or EEG signal. During severe hypoxia, for example, the EEG signal is already flat and therefore SD cannot depress it.}, is a slowly traveling wave (mm/min) characterized by neuronal depolarization and redistribution of ions between the intra- and extracellular space, that temporarily depresses electrical activity \citep{LEAO1947}, see Figure \ref{fig:KandV}. The phenomenon occurs in many neurological conditions, such as migraine with aura, ischemic stroke, traumatic brain injury and possibly epilepsy \citep{Dreier2011, Lauritzen2011}. Migraine is the most prevalent condition in which SD occurs and causes significant disability \citep{Leonardi2013}. SD seems to be relatively harmless for the neural tissue in the case of migraine aura, where a functional increase in blood flow enables a fast recovery. SD also occurs in ischemic stroke, where it can aggravate ischemic damage and its occurrence has been shown to correlate with poor outcome \citep{Nakamura2010, Hartings2011b}.

SD is a reaction-diffusion (RD) process, similar to the propagation of a flame on a matchstick \citep{Zeldovich1938}. 
SD consists of local ``reaction'' processes, such as release of potassium and glutamate, pump activity and recovery of the tissue in a later stage, as well as diffusion of potassium and glutamate, which enables the propagation of SD. Knowledge of the local dynamics and propagation of SD is essential for designing successful therapies that prevent or halt migraine attacks, or protect tissue in the penumbra from secondary damage after ischemic stroke.

\subsubsection*{Modeling cardiac arrhythmia serves as example}
Research in the last five decades, starting with the seminal work of \citet{WIE46}, has shown that cardiac
arrhythmias can be explained in terms of nonlinear RD wave dynamics in 2D (or 3D). Whole heart computer models of arrhythmia can predict what happens to the heart, and they led to the development of new medical strategies \citep{Trayanova2011}. On the cellular level, models of action potentials in cardiac cells also incorporate ion dynamics, for example, to model
cardiac  beat-to-beat variations and higher-order rhythms in ischemic ventricular muscle \citep{DIF85,DOK93,NOB01}. These developments could serve as a role model for SD modeling in migraine and stroke research, and inform us in particular which questions require what type of model.

\subsubsection*{Two types of models for SD}
\begin{figure*}
	\includegraphics[width=1.00\textwidth]{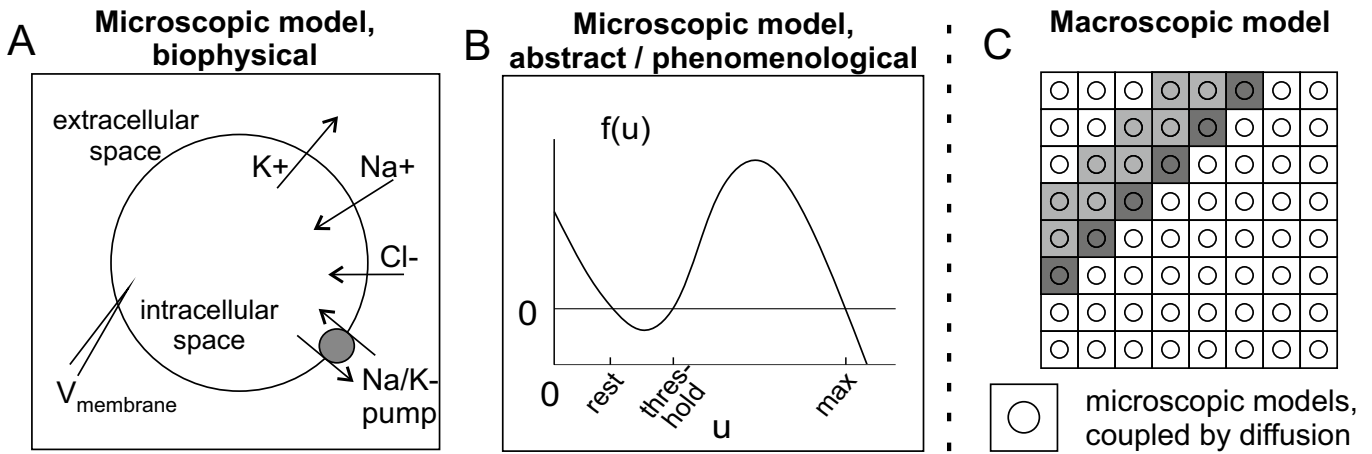}
	\caption{Modeling of SD. The microscopic interactions within the tissue (the ``reaction part'') can be modeled with a biophysical model, describing the ionic currents and release of neurotransmitters (panel A). Alternatively, a phenomenological or abstract model can be used, in which only one or two effective variables are considered that represent activation and recovery (panel B). Using these models, a plane or grid can be constructed to investigate SD propagation (Panel C).}
	\label{fig:modeltypes}
\end{figure*}

Computational models for SD conceptually consist of two parts: one part that models microscopic processes, i.e. the interactions within a single neurovascular unit leading to local failure of homeostasis and breakdown of the ion gradients, and a second part that describes the interactions throughout the tissue, usually through diffusion, leading to the macroscopic propagation of the homeostatic disturbance (Figure \ref{fig:modeltypes}). The latter is usually described by relatively simple expressions for diffusion. The microscopic interactions, however, are much more complex. For example, the concentration dynamics of potassium depend on the neuronal membrane voltage dynamics, buffering by glial cells and diffusion to the blood vessels. These microscopic processes can be modeled with either detailed biophysical models, or by more abstract models of so-called activator--inhibitor type.

The detailed biophysical models are suitable to investigate microscopic processes: the time-course of ions, transmitters, channels and pumps, and their contribution to SD. Abstract activator--inhibitor models are better suited to understand macroscopic behavior: the propagation and pattern formation of SD waves. However, many important questions include both aspects: How can non-invasive stimulation break up an SD wave? What is the neural correlate of EEG and fMRI signals recorded during peri-infarct depolarizations? Which combination of channel blockers efficiently blocks SD propagation? What are critical differences between human patients and animal models? These questions show the need for combining the two approaches, and linking parameters of abstract models to the behavior of biophysical models.

\subsubsection*{Outline}
This review discusses the two main types of models used to study SD, their advantages and disadvantages, and the recent advances in techniques to link them. We start however, by discussing the basic biophysics and physiology of SD that is used to construct these models.

\section{Physiology of SD}
The reviews of \citet{Somjen2001b} and \citet{Pietrobon2014} discuss the phenomenology, physiology and pharmacology of SD in great detail. Here we focus on the basic physiological and biophysical concepts important for computational modeling of SD.

Experimentally, SD can be induced by various stimuli, including ischemia, intense electrical stimulation, mechanical damage (needle prick) or application of K$^+$ or glutamate. These are all stimuli that directly or indirectly increase neuronal excitability or depolarize neuronal membranes. Similar to an action potential, once triggered, SD propagates in an all or none fashion, independent of the stimulus type or intensity.


Four hypotheses exist to explain the propagation of SD. The potassium and glutamate hypotheses state that SD propagates through diffusion of extracellular potassium or glutamate respectively. The neuronal gap junction hypothesis states that SD propagates by opening of neuronal gap junctions, while the glial hypothesis assumes that SD is caused by transmission through glial gap junctions. Evidence seems to favor the potassium hypothesis \citep{Pietrobon2014}, although neither of the hypotheses can fully explain the experimental observations, and propagation is probably realized by a combination of these mechanisms \citep{Somjen2001b}. In line with most modeling work on SD, we will also focus on release and diffusion of extracellular potassium and glutamate, and do not discuss propagation via neuronal or glial gap junctions.

First, we will discuss how extracellular potassium and glutamate stimulate their own release when homeostasis mechanisms are overchallenged and how this leads to sustained neuronal depolarization. Then, diffusion to neighboring tissue of the released substances is discussed and how movement of ions induces extracellular voltage gradients. Subsequently we elaborate on the recovery processes that enable restoration of the ion gradients and electrical activity and discuss the role of cell swelling and synapses in SD.

\subsection{Homeostasis of the neurovascular unit fails during SD}
Proper neuronal functioning relies on a steady supply of energy in the form of glucose and oxygen from the blood, as well as support from glia cells maintaining homeostasis of the extracellular composition. The neurovascular unit is a useful theoretical concept for describing (patho)physiology of neural metabolism. This unit consists of neuronal and glial intracellular space (ICS), the extracellular space (ECS) and a capillary supplying blood flow. The metabolic and homeostatic processes in such a unit determine largely how neural tissue reacts to ischemic and homeostatic insults such as SD and ischemia.

\begin{figure*}
	\centering
		\includegraphics[width=0.8\textwidth]{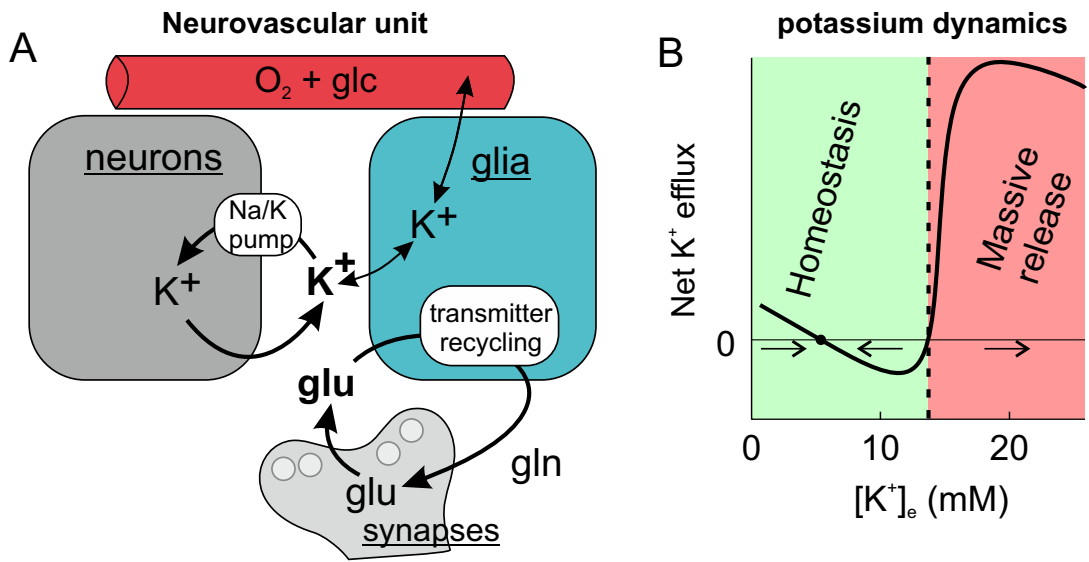}
	\caption{Homeostasis of extracellular potassium and glutamate in the neurovascular unit. Panel A schematically shows the main release and uptake pathways. Potassium leaking from the neurons and released during action potentials is pumped back by the Na/K-pump. Synaptically released glutamate (glu) is taken up by glia cells, and returned in the form of glutamine (gln). In addition, glia can rapidly buffer K$^+$ , distribute it over the glial syncytium and transport it to the blood stream. A constant supply of oxygen and glucose from the blood is necessary to fuel these processes. Adapted from \citep{Zandt_thesis}. Panel B shows a sketch of the dynamics of extracellular potassium. Up to a threshold (dashed line) of typically 8-20 mM, elevated extracellular potassium increases its own removal from the extracellular space by stimulating Na/K-pumps and glial uptake. This restores the concentration to the physiological set point (black dot). Above threshold (dashed line), potassium is released into the extracellular space faster than its removal due to stimulation of neuronal action potential generation. Based on \citep{Zandt2013}. The dynamics of extracellular glutamate show a similar threshold (not shown).}
	\label{fig:neurovascularunit}
\end{figure*}

Extracellular potassium and glutamate concentrations are tightly regulated in the brain. During rest, potassium ions leak from neurons, while action potentials and synaptic input increase this efflux even more. An estimated $70\%$ of the energy produced in the brain is consumed by neuronal Na/K-pumps and other ion transporters, in order to maintain physiological ion gradients over the neuronal membranes \citep{Howarth2012}. High extracellular potassium concentrations strongly increase neuronal excitability and hence glia cells rapidly take up excess amounts of potassium from the ECS. Furthermore, glia cells absorb glutamate released from excitatory synapses. Figure \ref{fig:neurovascularunit}A shows the main processes involved in the homeostasis of extracellular potassium and glutamate.

Rapid buffering of these two substances is critical, since they excite neurons and thereby stimulate their own release. This results in a positive feedback loop. Indeed, when a stimulus increases their concentration beyond a certain threshold, neuronal and glial transporters cannot cope with the efflux (Figure \ref{fig:neurovascularunit}B). This results in massive release of potassium and glutamate and leveling of the ion gradients, which disables the generation of action potentials.

\subsection{Sustained depolarization results from shifts in ion concentrations}
\label{sec:depolarization}

Each ionic species has a Nernst, or reversal, potential $E$ that drives the ionic current through the neuronal membrane. This electrical potential results from the concentration gradients across the semi-permeable membrane. Importantly, this voltage is determined by the intra- and extracellular ion concentrations:
\begin{equation}
\label{eq:Nernst}
E = \frac{RT}{zF} \textrm{ln}{\frac{C_{in}}{C_\text{out}}},
\end{equation}
where $F$ and $R$ are the Faraday and universal gas constant, $T$ the absolute temperature, and $z$ the valence of the ion species.
Although more accurate expressions, such as the Goldman-Hodgkin-Katz (GHK) equations, have been derived \citep{Keyner_and_Sneyd}, the Ohmic currents in the Hodgkin-Huxley (HH) equations suffice to qualitatively explain the neuronal electrophysiology during SD \citep{HUE14}. These show the resting membrane voltage is determined by the average Nernst potential, weighted by the respective ionic conductances $g$:
\begin{equation}
\label{eq:Vrest}
V_r = \frac{g_\text{Na} E_\text{Na}+g_\text{K} E_\text{K}+g_\text{Cl} E_\text{Cl}}{g_\text{Na} +g_\text{K} +g_\text{Cl}}.
\end{equation}
Hence, the neuronal membrane can be depolarized in two ways: changes in conductances and changes in Nernst potentials. An increased conductance of an outward current occurs for example during action potentials. During the upstroke of the action potential, the sudden opening of sodium channels temporarily generates an outward current that is not balanced by inward currents. This results in a fast (submillisecond) depolarization of the membrane voltage. The surplus of charge entering the cell resides in a very small region near the cell membrane \citep{Plonsey2007}, thus preserving electroneutrality in the solute. During SD, the glutamate level in the ECS rises \citep{Somjen2001b}, increasing the sodium conductance. This may induce the initial depolarization of neurons, according to the glutamate hypothesis.

\begin{table}[htbp]
  \centering
  \caption{{\bf Typical neuronal ion concentrations \citep{VanPutten2009} and corresponding Nernst potentials.}}
    \begin{tabular}{lccc}
              & Intracellular & Extracellular & Nernst Potential\\
              & (mM)					& (mM)					& (mV)\\
    Na$^+$    & 13 						& 140   				& 60 \\
    K$^+$     & 140 					& 4     				& -95 \\
    Cl$^-$    & 4 					  & 120   				& -90 \\
    \end{tabular}
  \label{tab:ionconc}
\end{table}

In contrast, the sustained depolarization and slow membrane voltage dynamics observed during SD are due to a more gradual (seconds) change of the resting membrane voltage, mediated by changing intra- and extracellular ion concentrations (equations \ref{eq:Nernst} and \ref{eq:Vrest}). Large numbers of ions flow across the membrane during SD, and these tend to equilibrate the concentrations of the ICS with the ECS. The membrane voltage and ion concentrations shift towards the Donnan equilibrium \citep{Somjen2001b}. In this equilibrium the ion gradients and membrane voltage are close to zero, but do not completely vanish due to large charged molecules in the ICS that cannot cross the neuronal membrane. The currents generated by ion pumps and transporters, as well as the slow Cl- dynamics, keep the cell from fully reaching the Donnan equilibrium. We will therefore refer to this depolarized state as the near-Donnan state, to distinguish it from ``ordinary'', conductance mediated, depolarization.

Extracellular potassium plays an important role in the triggering and propagation of SD. Of the main ionic species in the ECS and ICS, i.e. Na$^+$, K$^+$ and Cl$^-$, extracellular potassium influences the resting membrane potential most. Its concentration is relatively low (table \ref{tab:ionconc}) and the extracellular space relatively small. Hence, transmembrane fluxes can elevate this concentration relatively rapidly. In addition, the potassium conductance is relatively large such that $E_\text{rest}$ is close to $E_K$.

Note that, since the ECS and ICS need to remain electroneutral, and the capacitance of the membrane is limited, no net electrical current can flow across the membrane on the time scales of seconds or longer. Therefore, changes in ion concentrations and sustained depolarization cannot result from a single ionic current, but is rather mediated by a set of opposing currents. The necessity for balanced, opposing currents should be kept in mind when, for example, interpreting measurements in which specific currents are blocked to investigate which currents play a role in SD. For example, reducing the potassium conductance by partly blocking K$^+$ channels hardly lowers potassium efflux, since this is typically limited by the sodium influx. Instead, this depolarizes the resting membrane voltage \citep{Aitken1991} and when this depolarization is large enough, voltage gated sodium channels open, allowing for a rapid efflux of potassium and subsequent depolarization \citep{Kager2000, Zandt2011}.

\subsection{Diffusion of potassium and glutamate can propagate SD}
After potassium and glutamate are released locally, they diffuse to neighboring tissue, and can thereby propagate an SD (see Section \ref{sec:modeling}). During propagation, the front of the SD wave extends over several 100 $\mu$m's in the longitudinal direction, and hence SD propagation is a smooth process, rather than a chain reaction from neuron to neuron \citep{Yao2011}.

While diffusion from a fixed source becomes progressively slower over longer distances, an RD process (SD) propagates at a steady velocity by recruiting medium (tissue) at the front of the wave as new source. For an idealized case, the velocity is given as \citep{Zandt2013a}:
\begin{equation}
v = \sqrt{\frac{R D_\textrm{eff}}{\Delta C}},
\end{equation}
where $R$ is the rate at which neurons expulse potassium or glutamate, $D_{eff}$ the effective diffusion constant and $\Delta C$ the concentration threshold above which neurons start expulsing this substance.

\subsection*{Tortuosity}
The diffusion constants in water are $2.1 \times 10^{-9} \text{m}^2/\text{s}$  and $0.76 \times 10^{-9} \text{m}^2/\text{s}$ for K$^+$ and glutamate, respectively (at $25 ^o \textrm{ C}$) \citep{Bastug2005,Longsworth1953}. However, the large cell density in neural tissue hinders diffusion, and the effective diffusion coefficient $D_\text{eff}$ in ECS is typically a factor 2.5 lower than the free diffusion constant $D$. This is denoted by the tortuosity $\lambda$, historically defined as $\lambda^2 = D /D_\text{eff}$. Typically $\lambda = 1.6$ for ECS \citep{Nicholson1981,Nicholson1998}. \citet{Sykova2008} extensively review the physiology of diffusion in the extracellular space.

\subsubsection*{Electro-diffusion}
Often overlooked, however, is that in contrast to electrically neutral particles, potassium and glutamate are charged substances that cannot diffuse freely. A displacement of e.g. K$^+$ ions induces a voltage gradient in the tissue. The resulting electrical force (drift) counteracts the diffusion. Hence, the amount of K$^+$ that diffuses will be substantial only when there is counter movement of cations or co-movement of anions. This phenomenon is referred to as electro-diffusion. The voltage induced by the diffusion of ions creates a liquid junction potential and can be calculated with the Goldman-Hodgkin-Katz (GHK) expressions \citep{Keyner_and_Sneyd,Hille2001}. (For expressions correctly taking the transmembrane currents into account see \citep{Mori2014}.) The main contributors in ECS are K$^+$, Na$^+$ and Cl$^-$. Considering only these species, the extracellular voltage due to diffusion between two points in close proximity is calculated as:
\begin{equation}
\Delta V = \frac{RT}{F} \textrm{ln} \left( \frac{D_\text{k} [\textrm{K}^+]_1 + D_\text{na} [\textrm{Na}^+]_1 + D_\text{cl} [\textrm{Cl}^-]_2}{D_\text{k} [\textrm{K}^+]_2 + D_\text{na} [\textrm{Na}^+]_2 + D_\text{cl} [\textrm{Cl}^-]_1} \right),
\label{eq:Vghk}
\end{equation}
where the subscripts 1 and 2 denotes the concentrations at the two points in the extracellular space. The extracellular currents are calculated for each ion species as \citep{Almeida2004}:
\begin{equation}
\vec{I} = \underbrace{-z F \frac{D}{\lambda^2} \vec{\nabla}C}_{\textrm{diffusion}} + \underbrace{\frac{z^2 F^2}{RT} \frac{D}{\lambda^2} C \vec{\nabla}V}_{\textrm{drift}},
\label{eq:Ighk}
\end{equation}
where C denotes the extracellular concentration of the ionic species and $z$ its valency. This expression was used by \citet{Qian1989} to adapt the cable equations for non-homogeneous ion concentrations in the ECS.

Using a numerical model including electro-diffusion in the ECS, \citet{Almeida2004} calculated the extracellular voltage during SD that arises from diffusion of K$^+$, Na$^+$ and Cl$^-$ to be approximately -14 mV, which is in agreement with experimental observations (cf. Figure \ref{fig:KandV}).


In most modeling studies of SD, extracellular potassium and glutamate are assumed to follow ordinary diffusion laws rather than those of electro-diffusion. This is a reasonable approximation, as long as a composite diffusion coefficient is used \citep{Helfferich1958}, which takes co- and counter-diffusion of the ions in the ECS into account.

\subsection{Recovery mechanisms}
\label{sec:recovery}
Under normoxic conditions, ion concentrations start to recover typically a minute after SD onset. Electrical activity returns after a few minutes. Several mechanisms contribute to the tissue's recovery. A critical factor is that the Na/K-pump has to overcome the potassium efflux. Therefore, mechanisms are necessary that reduce potassium efflux, stimulate pump activity, and support this activity by a sufficient supply of energy.

\subsubsection*{Na/K-pump and glial potassium removal}
To recover neuronal function, physiological ion concentrations in the ECS and ICS need to be restored after SD. Both increased intracellular sodium and extracellular potassium levels stimulate the Na/K-pump \citep{Somjen2004, Glitsch2001, Skou1974}. This is insufficient to counteract the potassium efflux, however. In fact, this insufficiency was what instigated the depolarization process in the first place. Therefore, a critical step in the recovery process is the repolarization of the neuronal membrane voltage. This closes the voltage gated channels, greatly diminishing the potassium efflux, thereby allowing the pump to restore the physiological concentrations. The repolarization is effected by glial buffering of extracellular potassium from the extracellular space \citep{Kager2000,HUE14a}, lowering $E_k$, and thereby the membrane voltage (equations \ref{eq:Nernst} and \ref{eq:Vrest}).

Depending on the type of cell and brain area, the transmembrane voltage can be near 0 mV during SD, at which transient and NMDA-gated sodium channels are inactivated. Therefore, a yet unidentified conductance is argued to be activated in these cells during SD \citep{Makarova2007}. The sodium current through this conductance delays the recovery process.


\subsubsection*{Functional hyperemia}
The increased activity of the Na/K-pumps must be met with an increased blood flow, i.e. functional hyperemia, supplying additional oxygen and glucose to the tissue. The signaling pathways for vasodilation following increased neural activity\footnote{``Neural activity'' is an ill-defined term. In neurovascular coupling, the term encompasses activity of the synapses, mitochondria and generation of action potentials.} are mediated by astrocytes \citep{Zonta2003,Anderson2003b}, and include amongst others Ca$^{2+}$, K$^+$, adenosine, nitric oxide and arachidonic acid \citep{Iadecola2007}. These pathways mainly sense neuronal activity, rather than oxygen or glucose availability \citep{Mangia2009f}. For large disturbances, such as SD, the vessel response is strongly non-linear. While moderate increases of extracellular potassium cause vasodilation, stronger increases induce vasoconstriction \citep[and references therein]{Farr2011}. The neurovascular response to SD typically shows a triphasic response (constriction, dilation, followed by a prolonged, slight constriction), but differs greatly over species and conditions, ranging from pure constriction to pure dilation \citep{Ayata2013}.

Neurovascular coupling is a subject of active investigation, mostly in the light of the blood-oxygen-level-dependent (BOLD) response recorded by functional MRI (fMRI) \citep{Drake2007, Sotero2008}. When investigating (hypoxic) SD, it should be kept in mind that the normal neurovascular response is altered by effects induced by SD and hypoxia, such as changes in pH \citep{Somjen2001b,Pearce1995,Dirnagl1990}.

\subsubsection*{Synaptic failure} SD induces temporary synaptic failure. This failure reduces synaptic currents and suppresses electrical activity, thereby reducing the neuronal energetic needs. The cause of the failure is presynaptic, evidenced by the facts that electrical activity remains suppressed for several minutes after repolarization and that neurons do generate action potentials upon application of glutamate during this period \citep{Lindquist2012}. Synaptic failure is induced by high extracellular levels of adenosine, a break-down product of ATP, preventing the vesicular release of glutamate. Adenosine levels may increase as a result of increased ATP consumption, as well as from the release of ATP in the ECS \citep{Lindquist2012, Schock2007}.

\subsection{Role of cell swelling and synaptic interactions in SD}
\subsubsection*{Cell swelling}
Neurons regulate their volume and intracellular osmotic values with a variety of ion transporters and exchangers, aided by stretch sensitive ion channels \citep{Churchwell1996}. Changes in ion concentrations during SD alter the osmolalities of the ECS and ICS. This induces osmotic influx of water and consequent cell swelling, thereby equalizing the osmolalities. Cell membranes are highly permeable to water and do not sustain significant osmotic pressures, such that water influx must fully equalize the osmotic values of the ICS and ECS \citep{Hoffmann2009k}.
Note that exchange of Na$^+$ and K$^+$ does not change osmotic values. Hence transmembrane fluxes of anions or divalent cations, e.g. Cl$^-$ or Ca$^2+$, are necessary for cell swelling to occur \citep{Mueller2000}. 

Most biophysical models of SD, discussed in section \ref{sec:conductancebased}, calculate the evolution of the ion concentrations. These models can therefore naturally be extended with cell swelling. With the notable exception of the model by \citet{Shapiro2001}, most computational work shows that cell swelling mainly follows the dynamics of the ion concentrations during SD, rather than having a fundamental role in the initiation and propagation.

\subsubsection*{Synapses}
\label{sec:synapses}
Synaptic transmission is not necessary for SD propagation \citep{Somjen2001b}, and perhaps therefore, current computational models for SD are restricted to neurons without synaptic input. This is certainly realistic in hypoxic conditions, where synapses quickly fail \citep{Hofmeijer2012}. In normoxic conditions, however, synapses function normally at the onset of SD. Therefore, neuronal activity is determined by network dynamics and inhibitory feedback, rather than by single cell dynamics alone. Since inhibitory neurons are also excited by elevated extracellular concentrations of potassium and glutamate, the corresponding increase of overall firing rates, and hence release of K$^+$ and glutamate, may be less drastic than for isolated cells. In correspondence, blocking (inhibitory) GABA receptors has been shown to induce SD \citep{Hablitz1989,Dreier2012f}. Furthermore, prodromals, intense neuronal firing before depolarization, may alter the neuronal activity around the front of the wave through long range synaptic connections, influencing SD propagation.

So far, little theoretical work has been performed on the influence of network activity, local inhibition and long range connections on SD propagation and initiation.


\section{Modeling spreading depression}
\label{sec:modeling}

Broadly speaking, two types of computational/mathematical models for SD and peri-infarct depolarizations can be distinguished.\footnote{For completeness, we mention the cellular automata model of \citet{Reshodko1975}, who model tissue as a grid of machines, that turn on and off depending on the state of their neighbors. The propagation of activity in such a grid is very similar to the propagation of reaction-diffusion waves. The automata are a computationally efficient method for numerically investigating propagation of RD waves in complex geometries. Due to the lack of congruence with physiological mechanisms of SD, this approach has recently seen little interest in SD research.}

On the one hand, there are bottom--up, biophysical models, whose variables describe physiological quantities. These models consist of sets of differential equations describing the neuronal membrane voltage dynamics, ion and neurotransmitter fluxes and concentrations, and activity of homeostasis mechanisms. These models extend the traditional conductance based, i.e., HH-type, models with dynamics of the concentrations of ions and neurotransmitters in the ECS and ICS. They typically contain several equations and many parameter values for conductances and pump rates.

On the other hand, there are more phenomenological or abstract models. These typically describe only the dynamics of one variable, e.g. the extracellular potassium concentration or the general level of excitation. Some models then add a second variable that summarizes the processes enabling recovery.

A bottom--up biophysical model is necessary if one is interested in the profile of various physiological quantities that play a role in the spread of SD. Since the equations represent clear biophysical interactions, the mode of action of e.g. a neuroprotective agent, channel mutation or stimulation can be included in such a model in a straightforward way. However, analyzing the dynamics of such a detailed model is complicated, and the investigator is left with performing experiments with the model in a similar manner as with real world tissue.

In contrast to the questions on the biophysics, there are clinical questions that concern the general propagation of SD. For example, the spread of SD waves in a full--scale
migraine attack with aura determines the sequence of various symptoms \citep{VIN07}. The SD pattern spans over tens of centimeters in the cortex and can last hours. In this case, there clearly is a need for a model of SD with simplified dynamics that effectively describes large and sustained patterns in 2D, without the need to follow all physiological quantities on a cellular level.

While such a more phenomenological or abstract model allows for mathematical analysis, revealing the basic properties of SD initiation and propagation, it is no longer possible to explicitly include the action of drugs or channel mutations in the model. Explicitly linking detailed and phenomenological models, i.e., deriving simplified models from more detailed ones, allows to investigate how such conditions affect the parameters of a simplified model.




\subsection{Conductance based models with dynamic ion concentrations}
\label{sec:conductancebased}
Several microscopic models have been constructed to describe the ionic fluxes/currents and corresponding dynamics of the concentrations in the intra- and extracellular spaces (Figure \ref{fig:modeltypes}A). Some of these models were specifically designed to describe neuronal depolarization and spreading depression, while others were designed to explain bursting and epileptiform activity induced by ion concentration dynamics. The latter can be used to investigate neuronal depolarization as well. The review of \citet{Miura2013} discusses the most prominently used models for SD in more detail, as well as the differences between these models and their specific findings. Here we will focus on the general form and use of these models.

\subsubsection*{Microscopic, single unit models}
The simplest current based models consider an extracellular space and a neuron modeled as a single (somatic) compartment \citep{Barreto2011,Cressman2009,Cressman_err,Zandt2011}. This compartment has a neuronal membrane with leak currents and voltage gated Na and K-channels as in the HH model, and a Na/K-pump. The ion concentration dynamics in the intracellular compartment are driven by the fluxes of ions through the neuronal membrane, i.e., the leak, gated and pump currents. The concentrations in the extracellular space are additionally regulated by homeostatic mechanisms such as diffusion to the blood and glial potassium buffering. The original HH model, with only two gated channels, was shown to be sufficient to explain the various types of membrane voltage dynamics observed during depolarization of rat pyramidal cells in vitro \citep{Zandt2013a}.

More detailed models have been constructed that include one or multiple dendritic compartments and/or additional ion channels \citep{Kager2000,Kager2002,Kager2007,Makarova2007,Bazhenov2008,Florence2009,Somjen2009,Krishnan2011,Oeyehaug2012}. These more elaborate models allow for better quantitative agreement with experimental data, and investigation of the contribution of specific ion channels to SD vulnerability and seizures.

The observed dynamics of these models are qualitatively all similar. In general, depolarization can be induced in these models by application of extracellular potassium or glutamate, release of potassium from intense stimulation or temporary halt of the Na/K-pump. This results in an initial moderate depolarization of the resting membrane voltage. If this depolarization is large enough, voltage gated sodium channels open, greatly increasing potassium efflux, leading to sustained depolarization.

These single unit models can be used to investigate what mechanisms trigger or prevent depolarization locally in the tissue, as well as the mechanisms for recovery. The more simple models allow for bifurcation analysis of the local ion dynamics, which can identify parameters, e.g. pump strengths or potassium inflow, that cause critical transitions between the physiological stable state, cycles of depolarization and recovery, or permanent depolarization \citep{Barreto2011, HUE14,HUE14a}.

\subsubsection*{Models with one- and two-dimensional space}
In order to investigate the actual propagation of SD, the above discussed microscopic models must be extended with a spatial component and extracellular diffusion (Figure \ref{fig:modeltypes}C) \citep{Tuckwell1978, Tuckwell1980,Shapiro2001,Yao2011}, or electro-diffusion \citep{Teixeira2004, Shapiro2001}.

Most models that investigate SD have at most two dimensions, since the cortex is basically a folded, two-dimensional, sheet. However, investigating propagation analytically is much simpler in one dimension. Therefore, models are often reduced to one dimension, by arguing that the wave front of SD is relatively straight. To increase computational speed and lower complexity further, some models neglect the dynamics of the voltage gated channels and thereby remove neuronal action potentials from the model \citep{Tuckwell1978, Tuckwell1980, Dronne2006}. This is justified because this simplification does not qualitatively alter the ion concentration dynamics during SD, although it may quantitatively alter, for example, the critical stimulus strength for inducing depolarization.


In addition to electro-diffusion, \citet{Shapiro2001} included gap junctions and cell swelling in his model. In support of the gap junction hypothesis, he finds that both the current through the gap junctions as well as the concentration increase due to cell swelling is necessary for SD propagation, while extracellular diffusion does not significantly contribute to SD propagation. He shows that this effect is robust for variation in the parameters. However, most other models produce propagating SD waves without gap junctions or cell swelling. A reason for this may be differences in the conductance parameters, which can change over orders of magnitude between cell types and brain areas. However, the exact reasons for the different findings are not clear, since these models are hard to analyze without further simplification. This illustrates the main drawback of such very detailed models.

Finally we remark that all current models used to investigate SD propagation are essentially isolated cell models, i.e., they do not include the effects of synaptic interactions and network dynamics on the ion concentration dynamics. \citet{Ullah2009} model the activity of a network of excitatory pyramidal cells and inhibitory interneurons and the corresponding ion concentration dynamics, although they do not explicitly consider SD. Their model would be suitable for investigating how inhibitory feedback and network dynamics affect triggering and propagation of SD, an issue on which research has been lacking so far.
%


\subsection{RD models of activator--inhibitor type}
\label{sec:RDtypemodels}

There is more literature on very basic reaction--diffusion (RD) models in SD
than we can cover in detail in this review
\citep[see][]{Tuckwell1978,Tuckwell1981,Almeida2004,Postnov2012}. We focus on SD pattern formation in three essential
steps from (i) modeling propagation of the wave front, to (ii) modeling propagation and recovery (a pulse) to (iii) modeling
localized patterns.

\subsubsection*{Wave front propagation}
\label{sec:RD-front}

\citet{GRA63} originally proposed the potassium hypothesis and---based on a
suggestion by Hodgkin that included mathematical analysis from Huxley---she was the first to present an RD model of extracellular potassium concentration ([K$^+$]$_e$) dynamics in neural tissue that supports her experimental
observations and leads to roughly the correct speed of SD \citep{BUR74}.

Grafstein considered the effects of potassium release by the cells and potassium removal by the blood flow.
The RD model describes the dynamics of the extracellular potassium concentration, [K$^+$]$_e$ or simply $u$, with a rate function $f(u)$ that is a third order polynomial
with roots $f(u)\equiv0$ chosen at resting level concentration, threshold concentration (later called ceiling level by
Heinemann and Lux \cite{HEI77}) and maximum concentration (cf. Figure \ref{fig:neurovascularunit}B). Together with diffusion this yields:\begin{eqnarray} 
 \frac{\partial u}{\partial t} &=&  f(u) + D_u \nabla^2u\,.   \label{eq:GHmodel}     
\end{eqnarray}

Since recovery is not modeled, [K$^+$]$_e$ is locally bistable 
and can be resting at either the physiological resting level or the maximum 
concentration (pathological state). The variable $u$, the  [K+]$_e$, 
is also called an {\it activator}, because it activates a positive feedback loop 
when above a certain threshold. A stimulus, i.e., local application of potassium, 
can increase [K$^+$]$_e$ above threshold, releasing additional K$^+$. 
A sufficiently large stimulus \citep{Idris2008} triggers a traveling wave front, i.e., an SD, that eventually recruits all the medium in its state. %


\subsubsection*{Including recovery - pulse propagation}
\label{sec:RD-pulse}

\citet{REG94,REG96} have built the first computational model that aimed at
reproducing the typical zigzag of a fortification pattern experienced as visual
field defects during migraine with aura, 
\citep{LAS41,RIC71}.
One part of this model is an RD model for SD based on potassium dynamics, similar
to that of Grafstein and Hodgkin. However, it introduced two new features.

First their model includes a second variable describing the recovery process that drives the maximum [K$^+$]$_e$ back to the physiological resting level. For uniformity we refer to this recovery variable as $v$ ($r$ in the original papers). This was not the first such model with recovery, see e.g. \citep{Tuckwell1978,Tuckwell1981}, but we emphasize it here, because it directly links with earlier and later models discussed in the previous and the next section.


The recovery process is modeled phenomenologically as an additional removal of potassium. This recovery process, described by $v$, is slowly activated when [K$^+$$_e$] increases:
\begin{eqnarray}
 \frac{\partial u}{\partial t} &=&  f(u) - v + D_u \nabla^2u\,,\label{eq:u} \\
 \frac{\partial v}{\partial t} &=&  \varepsilon (c_1 u - v)\, .    \label{eq:v}
\end{eqnarray} 
$c_1$ determines the magnitude and $\varepsilon$ the activation time of the recovery, which is on a slower time scale (minutes) than the potassium concentration dynamics. When $v$ becomes sufficiently large, [K$^+$]$_e$ recovers to the physiological resting level. After this recovery, $v$ remains heightened for some time, leading to absolute and relative refractory periods for inducing a second SD.

$v$ is also called an {\it inhibitor} as it inhibits the release of potassium ions (the {\it activator}). RD models of activator--inhibitor type account for many important types of pattern
formation, such as spiral--shaped waves. There is a vast body of literature of activator--inhibitor models on chemical waves and patterns \citep{KAP95a}, which directly applies to propagation of SD.

\subsubsection*{Global inhibition - localized patterns}
\label{sec:RD-localized}
The models discussed so far, cannot account for the observation, from noninvasive imaging and reported visual field defects, that SD waves in migraine aura may propagate as a spatially localized pattern within the two-dimensional (2D) cortical sheet, rather than engulfing the entire cortex \citep{DAH04b,DAH08d,CHA13a}. This localization in 2D requires a third mechanism (the first two being the activator for front propagation and the inhibitor for pulse propagation in 1D, which can only explain engulfing ring pulses in 2D). 

In fact, another major new feature in the model by Reggia et al. \cite{REG94,REG96} is going half way the third and last step from fronts to pulses to localized patterns. Their RD model is coupled to a neural network, used to predict the visual field defects during migraine with aura. In brief, the mean firing rate of neurons at time $t$ in each ``cell'' (population of neurons) is represented by an activation level $a(t)$. They phenomenologically let [K$^+$]$_e$ modulate this activation level, such that subthreshold increases of [K$^+$]$_e$ stimulated activity, while superthreshold concentrations depressed activity. Furthermore, the neural network has lateral synaptic connections, such that cortical cells excite nearby cells and inhibit cells more distant (``Mexican hat'' connectivity).

This is novel, because the model incorporates as spatial lateral coupling not only local diffusion, but also synaptic long--range connections. However, the neural network dynamics were not fed back to the actual RD model (Equation~(\ref{eq:v})) and hence it remains an open question how the local and long range synaptic connections in the neural network influence the SD dynamics (see Section \ref{sec:synapses}). Nevertheless, long--range coupling is an essential mechanism for the emergence of localized patterns.

\citet{DAH12b} proposed that long--range coupling is established by the neuroprotective effect of increased blood flow induced by SD (Section \ref{sec:recovery}). In their model this increase in blood flow was assumed to be global, and its effect was phenomenologically incorporated as an inhibition process throughout the entire tissue. This global inhibitory feedback limits the spread of SD to traveling, localized spots on the (two--dimensional) cortical sheet, protecting it from a larger---possibly engulfing---recruitment into this pathological state. 

\begin{figure} \begin{center}
\includegraphics[width=\columnwidth]{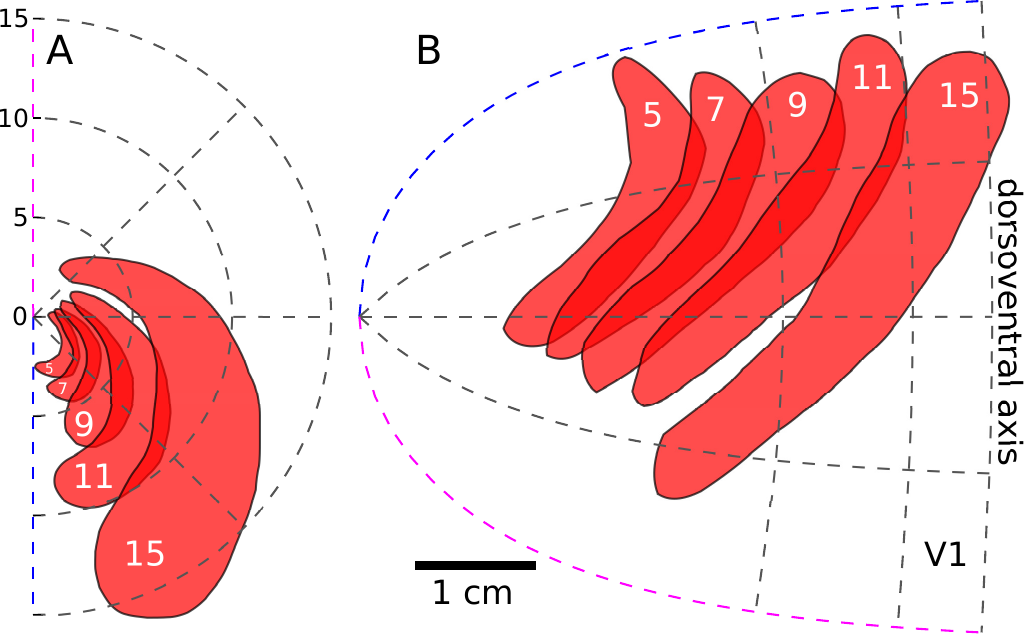} \end{center}
\caption{Five snapshots of a traveling visual migraine aura symptoms. From
the precise reports of visual symptoms by Lashley \cite{LAS41} of his own aura, for this particular
example the width of the wave front of 4 cm was estimated by retinotopically mapping the symptoms to
the primary visual cortex \citep{DAH08d}. \label{fig:las41Fig2Mapped}}
\end{figure}

According to the model, SD waves are initially spreading out radially. In a fraction
of simulated SD attacks, the circular wave breaks
open to a segment after not later than a few minutes
and then propagates further in one direction only. The arc length (width) of the SD front line, was estimated using this model to be between a
few millimeters up to several centimeters \citep{DAH12b}. This is in accordance with the precise reports of visual symptoms of his own aura by \citet{LAS41}, see
Fig.~\ref{fig:las41Fig2Mapped} left.

Furthermore, \citet{DAH12b} investigated the shape and form of SD patterns in single attacks, as well as their duration. They studied how these properties change for different degrees of cortical susceptibility to SD and claimed these emergent macroscopic properties can be linked to the prevalence of the major migraine subtypes, i.e., migraine with and without aura. They hypothesize that migraine pain induced by inflammation is only initiated if a large surface area is simultaneously covered by the SD pattern, and the aura symptoms, on the other side, can only be diagnosed if SD stays long enough ($>$5min) in the cortex. The analysis revealed that the severity of pain and aura duration are then to some degree anti-correlated and, furthermore, cortices being less susceptibility to SD can exhibit still short--lasting but significantly large SD patters that may underlay the concept of ``silent aura'', i.e., migraine without aura but pain caused by SD \citep{AYA10}.

\begin{figure}[t]
\begin{center}
\includegraphics[width=\columnwidth]{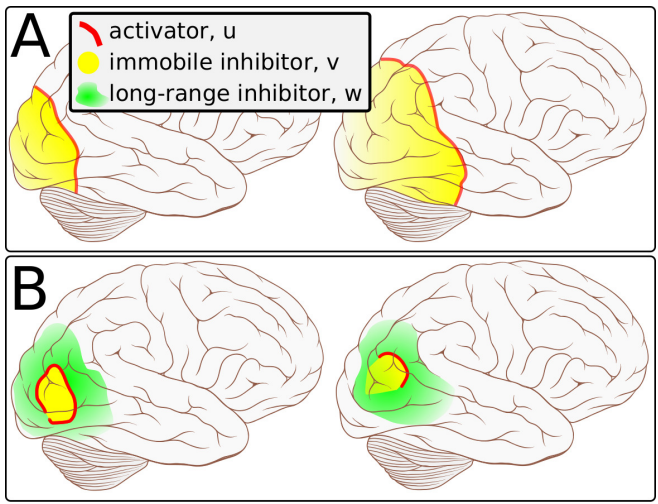}
\end{center}
\caption{{\bf Spatio-temporal development of SD.}  (A) Classical pattern formation paradigm, SD starting from an ictogenic focus in all directions and engulfing in a full--scale attack all of posterior cortex. (B) In the new paradigm, SD is a localized pattern that, when it breaks away from the ictogenic focus, it necessarily needs to break open and assumes the shape of a wave segment. Colors mark high activity (or concentration) of activator and inhibitors.}
\label{fig:csdEngulfingVsLocalized}
\end{figure}

The Dahlem model was inspired by a previous RD model \citep{KRI94} that shows propagation of spots in 2D
can be described by global inhibition. Such propagating spots were
observed in semiconductor material, gas discharge phenomena, and chemical
systems. Due to the global inhibition this is not a classical RD model. However, it closely
resembles a classical RD model with one activator $u$ and two inhibitors $v$ and
$w$ \citep{Woesler1996,Schenk1997}:
\begin{eqnarray} 
\frac{\partial u}{\partial t} &=&  f(u) - v - w + D_u \nabla^2u,   \label{eq:u2}  \\ 
\frac{\partial v}{\partial t} &=& \varepsilon \left(c_1 u - v\right)  , \label{eq:v3}\\ 
\frac{\partial w}{\partial t} &=& \theta \left(c_2 u - w \right) + D_w \nabla^2w .\label{eq:w}
\end{eqnarray}
When the diffusion constant $D_w$ is set very large, $w$ acts as global inhibitory feedback (see figure \ref{fig:csdEngulfingVsLocalized}).

The physiological substrate of these three lumped variables will be further elaborated on in the next
section. As such, these RD models are merely top--level descriptions that still lack a solid bottom--up derivation from the
positive feedback loop in potassium and other ion concentrations, electrical activity and homeostatic recovery mechanisms.

\subsection{Linking generic RD models to conductance based models}
\label{sec:linking}
\subsubsection*{Model reduction}
The ad hoc description in Sect.~\ref{sec:RDtypemodels} leaves important
questions unanswered: What exactly are these lump variables $u$, $v$, and
$w$? Which quantities have been lumped together and how? Is a polynomial rate function a generic description?
To answer these, one might think of adding more and more details to the top--level description started by
Grafstein \cite{GRA63} and eventually reach a description on the level of
conductance--based models with dynamic ion concentrations, but the reverse way
is more natural, a bottom--up approach starting from a conductance--based model with dynamic ion
concentrations. 

The first of two key steps is to reduce the conductance--based models. These models can contain several dozens of
dynamical variables. A reduction makes them tractable for a detailed
bifurcation analysis using a continuation software package, like AUTO \citep{DOE09}
that further leads to generic RD models with lump variables once the bifurcations have been identified.  

For example, several reduction techniques such as adiabatic elimination, synchronization, mass conservation, and
electroneutrality have been used to reduce a conductance based model of SD to only four dynamic
variables, while this model still retains adequate biophysical realism
\citep{HUE14}.

\subsubsection*{Towards identifying u, v, and w}
To further discuss the details of the activator $u$ and two yet unknown inhibitors
$v$ and $w$, it is insightful to divide the
reduced conductance--based model with dynamic ion concentrations into two
parts: the intracellular and extracellular compartment with the separating
membrane containing the voltage-gated channels (the cellular system)
and some ion buffer (the reservoir).

\begin{figure} \centering \includegraphics[width=0.5\textwidth]{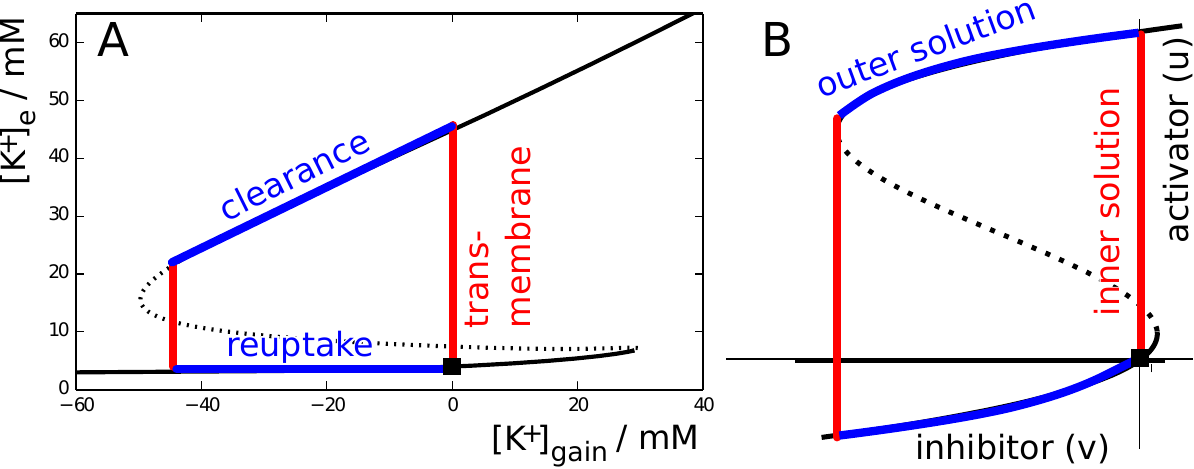}
  \caption{(A) Bifurcation diagram of HH model with dynamic ion concentrations using mols of potassium ions gained from a reservoir as a bifurcation parameter \citep{HUE14a}. Note that this parameter is given in terms of a concentration with the reference volume being that of the extracellular space. SD is marked as a counter-clockwise cyclic process starting from the physiological state (black square). Vertical subprocesses (red) occur at constant ion content, i.e., these are the fast pure transmembrane ion fluxes. Subprocesses with a horizontal component (blue) involve potassium ion clearance (right to left) or ion reuptake (left to right). The dotted and solid black line is the unstable and stable part, respectively, of the fixed point branch. The dotted line marks a threshold. (B) Phase space of a generic (i.e., polynomial) model described in Eqs.~(\ref{eq:u})-(\ref{eq:v}). Subsections of an excitation cycle can be obtained for two limits, the fast transitions (red) termed inner solution or threshold reduction, and the slow outer solutions (blue).} \label{fig:linkingRD}
\end{figure}

As discussed, the activator dynamics without an inhibitor (recovery mechanism) (Equation~(\ref{eq:GHmodel})) lead to a bistability.
In a conductance based model, the isolated cellular system without coupling to a reservoir was also found to be bistable \citep{HUE14}.
The bistability is seen in extracellular [K$^+$]$_e$ (Figure~\ref{fig:linkingRD}), but it is best
characterized by the full state of the cellular system. One stable state is the 
physiological resting state, far from thermodynamic equilibrium. The other state is the depolarized, near-Donnan state, close to the thermodynamic equilibrium of a semipermeable membrane (section \ref{sec:depolarization}). These two states are separated by an unstable equilibrium. 

In this closed system, the activator $u$ can directly be identified as the amalgamation of the variables that form the bistability, notably the extracellular potassium concentration.


Following this ansatz further, the local inhibitor $v$ is identified as the
potassium ion gain via external reservoirs, i.e., blood and glia cells. A loss (negative gain) of potassium ions both renders the near-Donnan state unstable, causing the system to recover, as well as increases the threshold for SD in the physiological resting state.
This reservoir coupling takes place on the slowest time scale of
the system \citep{HUE14a}, and hence the potassium ion gain can also be
considered as a bifurcation parameter (Fig.\ \ref{fig:linkingRD}). Note that
the ion gain as a bifurcation parameter is qualitatively different from the ion
bath concentration in the reservoir, which is often used as a bifurcation
parameter \citep{Bazhenov2008,Florence2009,Barreto2011,Krishnan2011}. In fact,
the essential importance of the potassium ion gain as a slow inhibitor and
hence useful bifurcation parameter was not realized in earlier studies
\citep{HUE14a}.


The long--range inhibitor $w$, may be related to neuronal activation, i.e., action potentials, for example through changes in long--range synaptic activity. Alternatively, it may be related to global blood flow regulation. However, $w$ cannot be identified from a model of a single neurovascular unit, since it represents an interaction that is essentially non-local.

\subsubsection*{Bifurcation analysis and generic models}

The second key step is to link this reduced conductance based model to a
generic RD model with lump variables in a more rigorous way. To do this, the
bifurcations in the reduced model must be analyzed and generic models of these
bifurcation can be considered as qualitative descriptions of SD. A generic
rate function of the activator would for example be a cubic polynomial,
resulting in two stable states and a threshold (cf. Section\
\ref{sec:RDtypemodels}).

To obtain a generic rate function of the inhibitor, one needs to analyze the
reduced conductance based model when the ion gain via some reservoir is not
treated as a bifurcation parameter. Then the onset of a cyclic SD process is
caused by a subcritical Neimark--Sacker bifurcation from a state of tonic
firing \citep{HUE14a}. When neglecting the fast time scale of tonic firing, the
subcritical Neimark--Sacker reduces to a subcritical Hopf bifurcation,
corresponding to type II excitability (if we adopt the classification scheme
from action potentials to SD).  Hence, this justifies---in hindsight---the use
of models showing such type II excitability for SD, as described in Section
\ref{sec:RDtypemodels}. Note however, that the model Eqs.~(\ref{eq:u2})-(\ref{eq:v3})
shows type II excitability with a supercritical Hopf bifurcation and subsequent canard explosion.

While this is still work in progress, such reduction techniques open up a
systematic study of the bifurcation structure as a valuable diagnostic method
to understand activation and inhibition of a new excitability in ion
homeostasis which emerges in HH models with dynamic ion concentrations.  This provides the missing
link between the HH formalism and activator--inhibitor models that have been
successfully used for modeling peri-infarct depolarizations and migraine
phenotypes.

\section{Conclusions and Outlook}
Spreading depression is the substrate of the migraine aura, and enhances infarct growth into the penumbra in stroke. The complex interplay of neuronal, homeostatic and metabolic dynamics in SD and peri-infarct depolarizations hampers the interpretation of pharmacological experiments. We discussed the basic (patho)physiology and biophysics of SD. Since these are largely known, most open questions on SD \citep{Pietrobon2014} pertain the dynamics, interaction and relative contribution of the processes involved. In order to answer these, mathematical modeling is a useful and necessary tool. Many computational and mathematical models have been constructed, both biophysical and more abstract reaction--diffusion models, which have given insight in local ion dynamics, propagation mechanisms and pattern formation of SD.

Single cell conductance based models were discussed, which allow to study the local dynamics of intra- and extracellular ion concentrations and the neuronal membrane voltage during depolarization. In order to study SD propagation, a sheet of single cell models with extracellular spaces connected by diffusion can be constructed. While these are suitable for in silico experimentation, better insight in the mechanisms of initiation, propagation and pattern formation can be obtained by using general reaction-diffusion equations of activator--inhibitor type.

The RD activator--inhibitor models, in turn, have the disadvantage of not describing the underlying microscopic processes. To investigate how microscopic interactions of e.g. ion channels and drugs determine the occurrence and propagation of SD, conductance based single cell models can be reduced or linked to a general form that can be analyzed analytically. While it was discussed that in the HH model with dynamic ion concentrations the extracellular potassium concentration and potassium buffering are linked to respectively the activator and the inhibitor, a general method for making such a reduction is still work in progress. In addition, the long range interactions responsible for confining the spatial extent of SD, for which functional hyperemia and long range synaptic connections have been suggested as candidates, still need to be identified.

An important next step in modeling the spatial spread of SD in migraine is to include regional heterogeneity of the cerebral cortex. The cortex is not simply a 2D surface, but is a sheet with thickness variations and further areal, laminar, and cellular heterogeneity. The problem of SD spread in an individual person can therefore ultimately only be resolved in a neural tissue simulation that extends the requirements and constraints of circuit simulation methods on a cortical sheet by creating a tissue coordinate system that allows for geometrical analysis \citep{kozloski2011ultrascalable,dahlem2014hot}. Since the cortical heterogeneity is to some extent like a fingerprint an individual feature of each migraine sufferer, the goal in the future will be to upload patient's MRI scanner readings into neural tissue simulators that then can deliver the same output as clinical data. This can be used as a test bed for exploring the development of stereotactic neuromodulation.

A next step for modeling SD in ischemia, is analyzing both detailed \citep{Dronne2006,Dronne2008} and more phenomenological models \citep{Revett1998,Vatov2006, Chapuisat2010} that include energy availability, to identify the key factors that determine frequency and duration of SD's. Reducing SD after stroke and global ischemia is a potential target for therapy \citep{Lauritzen2011}. For example, patients with global ischemia (and trials in focal ischemia are ongoing \citep{Worp2010,clintrials}) can benefit from mild therapeutic hypothermia \citep{Nielsen2013}. A possible mechanism is a reduced occurrence of SD. Analyzing SD dynamics during ischemia can not only help in selecting potential targets for neuroprotective agents or therapies, but also clarify the corresponding time window for successful application and elucidate critical differences between animal stroke models and human patients. These are key factors for the development of new neuroprotective drugs \citep{Feuerstein2009}.

In conclusion, modeling allows to further analyze the complex, dynamical phenomenon that is SD and can thereby aid in developing new treatments for migraine and stroke patients.

\section{Acknowledgments}
The authors would like to thank the Fields Institute for hosting the workshop on Cortical Spreading Depression and Related Neurological Phenomena, allowing us to incorporate new recent insights in this review.


\end{document}